\documentclass{birkjour}

\usepackage{amsthm, amsmath, amssymb, amsfonts, graphicx, subfig, psfrag}

\usepackage{color, hyperref}
\hypersetup{pdfstartview={XYZ null null 1.5}, pdfpagemode=UseNone}

\newcommand{\mat}[1]{\left[\begin{array}{cc}#1\end{array}\right]}
 \newtheorem{thm}{Theorem}[section]
 
 \newtheorem{lem}[thm]{Lemma}
 
 \theoremstyle{definition}
 
 \theoremstyle{remark}
 \newtheorem{rem}[thm]{Remark}
 
 \numberwithin{equation}{section}

\newcommand{\field}[1]{\mathbb{#1}}

\newcommand{\F}{\field{F}}

\newcommand{\ideal}[1]{\langle {#1} \rangle}
\newcommand{\radsp}{\mbox{\rm radsp}}
\renewcommand{\mod}{\mbox{\rm mod~}}
\newcommand{\coeff}{\operatorname{coef}}
\newcommand{\abs}[1]{\lvert #1 \rvert}

\newcommand{\inparen }[1]{\left(#1\right)}             
\newcommand{\inbrace }[1]{\left\{#1\right\}}           
\newcommand{\insquare}[1]{\left[#1\right]}

\newcommand{\cJ}{\mathcal{J}}
\newcommand{\cH}{\mathcal{H}}
\newcommand{\cR}{\mathcal{R}}
\newcommand{\cP}{\mathcal{P}}

\def\vec#1{\overline{#1}}
\DeclareMathOperator{\ch}{char}
\DeclareMathOperator{\rk}{rk}
\DeclareMathOperator{\poly}{poly}
\DeclareMathOperator{\trdeg}{trdeg}
\DeclareMathOperator{\size}{size}
\DeclareMathOperator{\degr}{deg}

\begin{document}

%
%
%
%
%
%
%
%
%

\title[Progress on PIT]
 {Progress on Polynomial Identity Testing - II}

\author[Saxena]{Nitin Saxena}

\address{%
Department of CSE\\
IIT Kanpur\\
Kanpur 208016\\
India}

\email{nitin@cse.iitk.ac.in}




\dedicatory{To my grand-advisor Professor Somenath Biswas}

\begin{abstract}
We survey the area of algebraic complexity theory; with the focus being on the problem of polynomial identity testing (PIT). We discuss the key ideas that have gone into the results of the last few years.
\end{abstract}

\maketitle
\tableofcontents

\section{Introduction}
\addtocontents{toc}{\protect\setcounter{tocdepth}{1}}

Algebraic complexity theory is the study of computation via {\em algebraic} models, hence, algebraic techniques. In this article we work with only one model -- {\em arithmetic circuit} (in short, {\em circuit}).  A circuit $C(x_1,\ldots,x_n)$, over a ring $R$, computes a polynomial $f$ in $R[x_1,\ldots,x_n]$. Its description is in the form of a rooted tree; with the {\em leaves} inputting the variables or constants, the internal {\em nodes} computing addition or multiplication, and the {\em root} outputting the $f$. The edges in $C$, called {\em wires}, carry the intermediate polynomials and could also be used to multiply by a constant (from $R$). By the {\em size}, respectively the {\em depth}, of $C$ we mean the natural thing (sometimes to avoid ``trivialities'' we might want to take into account the bit-size needed to represent an element in $R$).

A moment's thought would suggest that a circuit is a rather compact way of representing polynomials. Eg.~a circuit of size $s$ could produce a polynomial of degree $2^s$ (hint: repeated squaring). In fact, a single product gate could multiply $s$ linear polynomials and produce $n^{\Omega(s)}$ many monomials. Thus, a circuit is an `exponentially' compact representation of some polynomial families. Conversely, are there `explicit' polynomial families (say $n$-variate $n$-degree) that require exponential (i.e.~$2^n$) sized circuits? We ``expect'' almost every polynomial to be this hard, but, the question of finding an {\em explicit} family is open and is the main goal motivating the development of algebraic complexity. 

One can try to directly give a good {\em lower bound} against circuits by designing an explicit polynomial family $\{f_n\}$ and prove that it requires a `large' sized circuit family $\{C_n\}$. The other, indirect, way is to design an efficient {\em hitting-set} $\cH$ for the circuit family, i.e.~if $C_n\ne0$ then $\exists a\in\cH$, $C_n(a)\ne0$. This `flip' from lower bounds to algorithms was first remarked by \cite{HS80} and now it has several improved versions \cite{KI04, Agr05, Agr06}. This is a remarkable phenomena and is one of the primary motivations to study the question of PIT: Given a circuit $C$ test it for zeroness, in time polynomial in size$(C)$. The hitting-set version of PIT is also called {\em blackbox} PIT (contrasted with {\em whitebox} PIT).

The last 10 years have seen a decent growth of algebraic tools and techniques to understand the properties of polynomials that a circuit computes. The feeling is that these polynomials are special, different from general polynomials, but a strong enough algebraic `invariant' or a combinatorial `concept' is still lacking. There have been several articles surveying the known techniques and the history of PIT \cite{Sax09, AS09, SY10, CKW11, Sap13}. In this survey we will attempt not to repeat what those surveys have already covered. So, we will focus only on the new ideas and assume that the reader has given at least a cursory glance at the older ones. We directly move on to the Leitfaden.

\subsection{Survey overview}

This article deals mainly with three broad topics -- the `universality' of  depth-$3$ circuits, the design of hitting-sets via `faithful' morphisms and that via rank `concentration'. A major emerging area that we skip in this article is that of PIT vis \`a vis GCT (geometric complexity theory) program \cite{Mul11, Mul12, Mul-cacm12}; the algebraic-geometry interpretations there are interesting though any concrete PIT algorithm, or application, is yet to emerge. 

\smallskip
\noindent{\bf Shallow circuits.} 
A depth-$2$ circuit (top $+$ gate) of size $s$, over a field, essentially computes a sum of $s$ monomials. Such polynomials are called {\em sparse} polynomials; blackbox PIT for them was solved few decades ago. So, our next stop is depth-$3$: Polynomials of the form $$C = \sum_{i=1}^k \prod_{j=1}^d L_{i,j}, $$ where $L_{i,j}$ are linear polynomials in $\F[x_1,\ldots,x_n]$. Significant research has been done with this model, but both sub-exponential PIT and exponential lower bounds are open here. Recently, a remarkable universality result was shown for depth-$3$ \cite{GKKS13}: If an $n$-variate $\poly(n)$-degree polynomial can be nontrivially computed by a circuit, then it can be nontrivially computed in depth-$3$. This `squashing' of depth means that it suffices to focus on depth-$3$ for PIT purposes.   

If we consider a depth-$2$ circuit (top $\times$ gate), over a {\em ring} $R$, then again we get some remarkable connections. Fix $R$ to be the $2\times2$ matrix algebra $M_2(\F)$, and consider the circuit $$D=\prod_{i=1}^d L_i,$$ where $L_i$ are linear polynomials in $R[x_1,\ldots,x_n]$. Traditionally, $D$ is called a {\em width-$2$ algebraic branching program} (ABP). It was shown by \cite{SSS09} that depth-$3$ PIT efficiently reduces to width-$2$ ABP PIT.

\smallskip
\noindent{\bf Faithful morphisms.}
It was observed in the last few years that in all the known hitting-sets, the key idea in the proof is to work with a {\em homomorphism} $\varphi$ and an algebraic {\em property} that the image of $\varphi$ should preserve. \cite{SS12} used a (Vandermonde-based) map $\varphi:\F[x_1,\ldots,x_n]\rightarrow\F[y_1,\ldots,y_k]$ that preserves the `linear' rank of any $k$ linear polynomials. This gave the first blackbox PIT for bounded top fanin depth-$3$, over any field.  

\cite{BMS13, ASSS12} used a (Vandermonde \& Kronecker-based) map $\varphi:\F[x_1,\ldots,x_n]\rightarrow\F[y_1,\ldots,y_k]$ that preserves the `algebraic' rank (formally, {\em transcendence degree}) of certain $k$ polynomials. This gave the first blackbox PIT (and lower bounds) for several well-studied classes of constant-depth circuits. One drawback of the technique is that it requires zero/large characteristic fields.

\smallskip
\noindent{\bf Rank concentration.}
Inspired from the tensors, a restricted circuit model called multilinear read-once ABP (ROABP) has been intensively studied. Let $R$ be the $w\times w$ matrix algebra $M_w(\F)$ and let $\{S_i\}$ be a partition of $[n]$. Consider the circuit $D=\prod_{i=1}^d L_i,$ where $L_i$ are linear polynomials in $R[x_{S_i}]$ (i.e.~the linear factors have disjoint variables). For $D$ \cite{FSS13} gave a hitting-set in time $\poly(wn)^{\log w\cdot\log n}$, i.e.~quasi-poly-time. The proof is based on the idea, following \cite{ASS13}, that after applying a small (Kronecker-based) `shift', $D$ gets the property: The rank of its coefficients (viewed as $\F$-vectors) is concentrated in the `low' support monomials. Thus, checking the zeroness of these low monomials is enough!

We conjecture that rank-concentration, after a `small' shift, should be attainable in any ABP $D$. But, currently the proof techniques are not that strong. Recently, \cite{AGKS13} have achieved rank-concentration in multilinear depth-$3$ circuits where the partitions (corresponding to each product gate) are `close' to each other in the sense of `refinement'.

\section{Shallow circuits, deep interconnections}

In this section we exhibit the key ideas behind the universality of two shallow circuits.

\subsection{The depth-$3$ chasm}

In the study of circuits one feels that low-depth should already hold the key. This feeling was confirmed in a series of work \cite{VSBR83, AV08, Koi12, Tav13}: Any $\poly(n)$-degree $n$-variate polynomial computed by a $\poly(n)$-sized circuit $C$ can also be computed by a $n^{O(\sqrt{n})}$ sized depth-$4$ circuit! 

The idea for this is, in retrospect, simple -- since the degree is only $\poly(n)$, first, squash the depth of $C$ to $O(\log n)$ by only a polynomial blowup in the size (the product gates we get are quite {\em balanced}). Next, identify a subcircuit $C_2$ by picking those gates whose output polynomial has degree at least $\sqrt{n}$, and call the remaining subcircuit $C_1$. We view $C_2$ as our circuit of interest that takes gates of $C_1$ as input. It can be shown that $C_2$ computes a polynomial of degree $\approx\sqrt{n}$ of its input variables (which are $\poly(n)$ many). Obviously, each gate of $C_1$ also computes a polynomial of degree $\approx\sqrt{n}$ of its input variables (which are $x_1,\ldots,x_n$). Thus, $C_2$ finally computes a sum of $\approx\binom{\poly(n)+\sqrt{n}}{\sqrt{n}}$ products, each product has $\sqrt{n}$ factors, and each factor is itself a sum of $\approx\binom{n+\sqrt{n}}{\sqrt{n}}$ degree-$\sqrt{n}$ monomials. To put it simply, $C$ can be expressed as a $\sum\prod^{\sqrt{n}}\sum\prod^{\sqrt{n}}$ circuit of size $n^{O(\sqrt{n})}$. The details of this proof can be seen in \cite{Tav13}.

The strength of depth-$4$ is surprising. Recently, an even more surprising reduction has been shown \cite{GKKS13} -- that to depth-$3$ (again, $n^{O(\sqrt{n})}$ sized). We will now sketch the proof; it ties together the known results in an unexpected way. 

Essentially, the idea is to modify a $\sum\prod^a\sum\prod^a$ circuit $C$ of size $s:=n^a$ (where $a:=\sqrt{n}$) by using two polynomial identities that are in a way ``inverse'' of each other, and are to do with powers-of-linear-forms. First, replace the product gates using Fischer's identity:

\begin{lem}[\cite{Fis94}]
Any degree $a$ monomial can be expressed as a linear combination of $2^{a-1}$ $a$-th powers of linear polynomials, as:
$$y_1\cdots y_a\ =\ (2^{a-1}\cdot a!)^{-1}\cdot\mathop{\sum}_{r_2,\ldots,r_a\in\{\pm1\}} \inparen{ y_1 + \mathop{\sum}_{i=2}^{a} r_i y_i }^a\cdot (-1)^{\#\{i|r_i=-1\}}. $$
\end{lem}

We denote this type of a circuit by the notation $\sum\bigwedge^a\sum$, where the wedge signifies the powering by $a$. The above identity transforms the $\sum\prod^a\sum\prod^a$ circuit $C$ to a $\sum\bigwedge^a\sum\bigwedge^a\sum$ circuit, of size $\approx s$.

Next, the two power gates are `opened' up using an identity introduced by the author:

\begin{lem}[\cite{Sax08}]\label{lem-sax08}
For any $a,m$, there exist degree-$a$ univariate polynomials $f_{i,j}$ such that
$$(y_1+\cdots+y_m)^a\ =\ \mathop{\sum}_{i=1}^{ma+1} \mathop{\prod}_{j=1}^{m} f_{i,j}(y_j).$$
\end{lem}

Let us carefully see the jugglery on $C$. The $\sum\bigwedge^a\sum\bigwedge^a\sum$ circuit $C$ has the expression $C=\sum_i T_i$, where each $T_i$ has the form $(\sum_{j=1}^s \ell_{i,j}^{e_{i,j}})^{a}$ with linear $\ell_{i,j}$'s. We want to open up the top power gate of $C$. By Lemma \ref{lem-sax08} we get
$$T_i\ =\ \mathop{\sum}_{u=1}^{sa+1} \mathop{\prod}_{j=1}^{s} f_{u,j}(\ell_{i,j}^{e_{i,j}}).$$

Since $f_{u,j}$ is a univariate, it splits into linear polynomials when the base field $\F$ is {\em algebraically closed}. As $\ell_{i,j}$ is already a linear polynomial, we deduce that $T_i$, and hence $C$, is a $\sum\prod\sum$ circuit of size $\poly(s)$.

Finally, note that for the above arguments to work we require $\F$ to be algebraically closed and char$(\F)>a$. Lemma \ref{lem-sax08} has been generalized to all fields by \cite{FS13a}, so it is likely that this depth-$3$ reduction can be extended to all fields.

The optimality of $n^{\sqrt{n}}$-size, in this reduction, is open. However, \cite{KSS13} showed that any decent improvement would lead to a proof of $VNP\ne VP$.

\subsection{The width-$2$ chasm}

Here we look at $\prod\sum$ circuits over a matrix algebra. Though the model $D=\prod_i L_i$, with linear $L_i\in R[x_1,\ldots,x_n]$, seems innocuous at first sight, a closer look proves the opposite! It can be shown fairly easily that: A polynomial computed by a constant-depth circuit (over a field) can as well be computed by a $D$ over a $3\times3$ matrix algebra \cite{BC88}. On the other extreme, by taking $R=M_n(\F)$ we can compute the {\em determinant} of a matrix in $\F^{n\times n}$ \cite{MV97}, hence, arithmetic {\em formulas} (not general circuits!) can be simulated in this model \cite{V79}.

Perhaps surprisingly, \cite{SSS09} showed that: A polynomial computed by a depth-$3$ circuit (over a field) can as well be computed by a $D$ over a $2\times2$ matrix algebra. This, togetherwith the previous subsection, makes the $\prod\sum$ circuits over $M_2(\F)$ quite strong. 

Say, we want to express the depth-$3$ circuit $C=\sum_{i=1}^k T_i$ in a $2\times2$ matrix product. Firstly, we express a product $T_i=\prod_{j=1}^d \ell_{i,j}$ as:
$$\mat{ \ell_{i,1} & 0\\ 0 & 1 } \cdots
\mat{ \ell_{i,d-1} & 0\\ 0 & 1 } \cdot 
\mat{ 1 & \ell_{i,d}\\ 0 & 1 }
=  
\mat{ T_i' & T_i\\ 0 & 1  },  \text{ where } T_i':=T_i/\ell_{i,d}.$$
  
Once we have such $k$ $2\times2$ matrices, each containing $T_i$ in the $(1,2)$-th place, we would like to sum the $T_i$'s in a `doubling' fashion (instead of one-by-one). 

We describe one step of the iteration. Let $\mat{ L_1 & L_2f\\ 0 & L_3  }$ \& $\mat{ M_1 & M_2g\\ 0 & M_3  }$ be encapsulating two intermediate summands $f$ and $g$. With the goal of getting (a multiple of) $f+g$ we consider the following, carefully designed, product:
$$\mat{ L_1 & L_2f\\ 0 & L_3  }\cdot 
\mat{ L_2M_3 & 0\\ 0 & L_1M_2 } \cdot 
\mat{ M_1 & M_2g\\ 0 & M_3  } $$
$$=\quad \mat{ L_1M_1L_2M_3 & L_2M_3L_1M_2(f+g)\\ 0 & L_3M_3L_1M_2 } $$  

After $\log k$ such iterations, we get a {\em multiple} of $C$ in the $(1,2)$-th entry of the final $2\times2$ matrix product. Note that the middle matrix, introduced in the LHS above, potentially doubles (in the degree of the entry polynomials) in each iteration. Thus, finally, $D$ is a product of $\poly(d2^{\log k})$ linear polynomials over $M_2(\F)$. Thus, the size blowup is only polynomial in going from depth-$3$ to width-$2$. 

\section{Faithful morphisms, hitting-sets}

In algebraic complexity the study of certain maps has been fruitful -- homomorphisms $\varphi:\cR:=\F[x_1,\ldots,x_n]\rightarrow\F[y_1,\ldots,y_k]=:\cR'$  such that the algebraic `relationship' of certain polynomials $\{f_1,\ldots,f_k\}$ does not change in the image of $\varphi$. When $f_i$'s are linear this boils down to a linear algebra question and we can easily design $\varphi$ in time $\poly(n)$ (hint: employ Vandermonde matrix). This business becomes complicated when $f_i$'s are non-linear. Then we have to ask how are $f_i$'s represented. If they are given via monomials 
then we invoke the Jacobian criterion to design $\varphi$, but the time complexity becomes exponential in $k$. Several variants of such faithful maps are discussed in the PhD thesis \cite{Mit13}. We sketch the ideas behind two basic maps here.

\subsection{Bounded fanin depth-3 blackbox PIT}

Let $C=\sum_{i\in[k]} T_i$ be a depth-$3$ circuit. When $k$ is constant, $C$ is naturally called {\em bounded fanin} depth-$3$. This case of PIT has, by now, a rich history \cite{DS07, KS07, KS11, SS11, KS09, SS13, SS12}. Several new techniques have sprung up from this model -- a locally decodable code structure, a rank-preserving map via extractors, Sylvester-Gallai configurations (higher-dimensions and all fields) and rank bounds. We will sketch here the main idea behind the poly-time blackbox PIT of bounded fanin depth-$3$. The details are quite technical and could be seen in \cite{SS13, SS12}.

\smallskip\noindent{\bf Vandermonde map.}
We define a homomorphism $\Psi_\beta$, for a $\beta\in\F$, as:
$$
\forall i\in[n],\ \ \Psi_\beta: x_i \mapsto \sum_{j=1}^{k}\beta^{ij}y_j, 
$$
and $\Psi_\beta(\alpha)=$ $\alpha$ for all $\alpha\in\F$. This (naturally) defines the action of $\Psi_\beta$, on {\em all} the elements of $\cR$, that preserves the ring operations. We have the following nice property, as a consequence of \cite[Lemma 6.1]{GR08}: 

\begin{lem}[$\Psi_\beta$ preserves $k$-rank]\label{lem-rk-preserv}
Let $S$ be a subset of linear forms in $\cR$ with $\rk(S)\le k$, and $|\F|>nk^2$. Then $\exists\beta\in\F,\, \rk(\psi_\beta(S)) = \rk(S)$.
\end{lem}

Intuitively, $\Psi_\beta$ is {\em faithful} to any algebraic object involving the elements in span$(S)$. The proof of this lemma is by studying the coefficient-matrix of the linear polynomials in $S$, and its change under $\Psi_\beta$. This map has a role to play in bounded fanin depth-$3$ owing to a certain structural theorem from \cite{SS13} -- {\em certificate for a non-identity}. 

To discuss this certificate we need a definition, that of `paths' of  `nodes' in $C$ (assumed to be nonzero). A {\em path} $\vec{p}$ with respect to an ideal $I$ is a sequence of terms $\{p_1, p_2, \ldots, p_b\}$ (these are products of linear forms) with the following property. Each $p_i$ divides $T_i$, and each $p_i$ is a `node' of $T_i$ with respect to the ideal $\ideal{I, p_1, p_2, \ldots, p_{i-1}}$.\footnote{By a {\em node} $p_i$ we mean that some nonzero constant multiple of $p_i$ is identical to a power-of-a-linear-form modulo $\radsp\ideal{I, p_1, p_2, \ldots, p_{i-1}}$, where $\radsp$ is the ideal generated by the set of all the linear polynomials that divide $p_j, j\in[i-1]$ and the generators of $I$.} So $p_1$ is a node of $T_1$ wrt $I$, $p_2$ is a node of $T_2$ wrt $\ideal{I,p_1}$, etc. 

\begin{figure}[tb!]
	\centering
	{ 
	\psfrag{T1}{$T_1$} \psfrag{T2}{$T_2$} \psfrag{T3}{$T_3$} 
	\psfrag{L1}{$x_1$} \psfrag{M1}{$x_1$} \psfrag{L2}{$x_2$} \psfrag{L3}{$x_2 + 2x_1$} 
	\psfrag{L4} {$x_3 + 10x_1$} \psfrag{L5} {$x_3 - x_1$} \psfrag{L6} {$x_3 + 3x_1$}
	\psfrag{L7} {$x_4 + x_2$} \psfrag{L8} {$x_4 +4x_2 - x_1$} \psfrag{L9} {$x_4 + x_2 + x_1$} \psfrag{L10} {$x_4 + x_2 - 2x_1$}
	\includegraphics[scale=0.4]{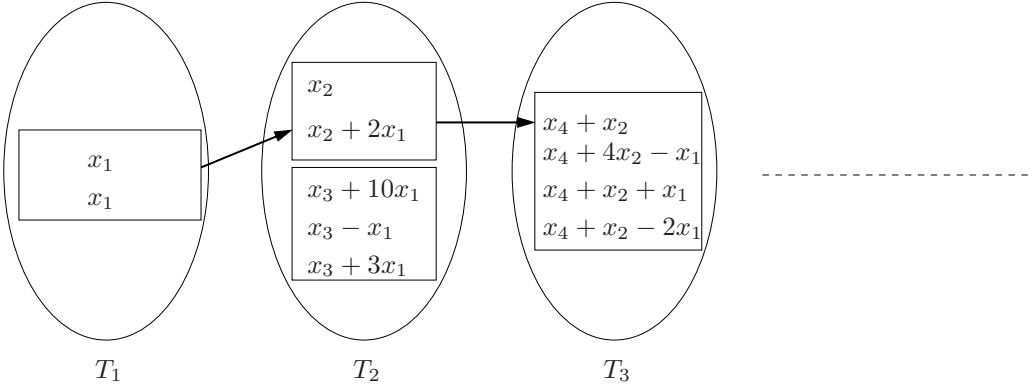}} 
	\caption{Nodes and paths in $C = T_1 + T_2 + T_3 + \ldots$} \label{fig-path}
\end{figure}

Let us see an example of a path $(\ideal{0}, p_1, p_2, p_3)$ in Figure~\ref{fig-path}. The oval bubbles represent the list of forms in a product gate, and the rectangles enclose forms in a node. The arrows show a path. Starting with the zero ideal, nodes $p_1 := x^2_1$, $p_2 := x_2(x_2+2x_1)$, and $p_3 := (x_4 + x_2)(x_4 +4x_2 - x_1)(x_4 + x_2 + x_1)(x_4 + x_2 - 2x_1)$ form a path. Initially the path is just the zero ideal, so $x^2_1$ is a node. Note how $p_2$ is a power of $x_2$ modulo $\radsp\ideal{p_1}$, and $p_3$ is a power of $x_4$ modulo $\radsp\ideal{p_1,p_2}$.

The non-identity certificate theorem \cite[Theorem 25]{SS13} states that for any non-identity $C$, there exists a path $\vec{p}$ such that modulo $\ideal{\vec{p}}$, $C$ reduces to a single nonzero multiplication term. 

\begin{thm}[Certificate for a non-identity]\label{thm-cert-non0-real}
Let $I$ be an ideal generated by some multiplication terms. Let $C=\sum_{i\in[k]}T_i$ be a depth-$3$ circuit that is nonzero modulo $I$. Then $\exists i\in\{0,\ldots,k-1\}$ such that 
$C_{[i]}$\footnote{We mean $C_{[i]}:=\sum_{j\in[i]} T_j$.} mod $I$ has a path $\vec{p}$ satisfying: 
$C\equiv \alpha\cdot T_{i+1}\not\equiv0$ $(\mod I+\ideal{\vec{p}})$ for some $\alpha\in\F^*$.
\end{thm}

The proof of this theorem involves an extension of Chinese remaindering to ideals that are generated by multiplication terms. Once we have this structural result about depth-$3$, observe that we would be done if we could somehow ensure $T_{i+1}\notin\ideal{\vec{p}}$ (in our application $I$ is zero). How do we preserve this ideal non-membership under a cheap map?

Notice that the rank of the set $S_0$ of linear polynomials that divide the nodes in the path $\vec{p}$ is $<k$ (since path length is below $k$). Moreover, $T_{i+1}$ factors into at most $d$ linear polynomials, denote  the set by $S_1$. So if we apply a map that preserves the rank of each of the $d$ sets $S_0\cup\{\ell\}, \ell\in S_1$, then, intuitively, the ideal non-membership should be preserved. As $\rk(S_0\cup\{\ell\})\le k$ we can employ the previously discussed map $\Psi_\beta$ (over a field satisfying $|\F|>dnk^2$). This idea could be easily turned into a proof; details are in \cite{SS12}.

Finally, what we have achieved is the construction of a map $\Psi_\beta$, in time $\poly(dnk)$, that reduces the variables of $C$ from $n$ to $k$ and preserves nonzeroness. Once this is done, the $\poly(nd^k)$ blackbox PIT follows from the brute-force hitting-set. 

\subsection{Depth$\ge3$ results}

Looking at the success of bounded fanin depth-$3$ one wonders about the analogous depth-$4$ model: 
\begin{equation}\label{eqn-depth4}
C\ =\ \mathop{\sum}_{i\in[k]}\mathop{\prod}_{j\in[d]}f_{i,j},\ \text{ where } f_{i,j} \text{ are sparse polynomials.}
\end{equation}
Here we are thinking of a bounded $k$. But now even $k=2$ seems nontrivial! In fact, a simpler PIT case than this is an old open question in a related area \cite{G83}.

This {\em bounded top fanin} depth-$4$ PIT is an important open question currently. What is doable are other restricted models of depth-$4$. Inspired from the last subsection we ask: Is there a notion of `rank' for general polynomials, are there easy `faithful' maps, and finally is all this useful in PIT?

There are several notions of rank in commutative algebra. The one we \cite{BMS13} found useful is -- {\em transcendence degree} (trdeg). We say that a set $S$ of polynomials $\{f_1,\ldots,f_m\}\subset\F[x_1,\ldots,x_n]$ is {\em algebraically dependent} if there exists a nonzero annihilating polynomial $A(y_1,\ldots,y_m)$, over $\F$, such that $A(f_1,\ldots,f_m)=0$. The largest number of algebraically {\em in}dependent polynomials in $S$ is called $\trdeg(S)$. With this notion we call a homomorphism $\varphi$ {\em faithful} if $\trdeg(S) = \trdeg(\varphi(S))$.
The usefulness of $\varphi$ (assuming that one can come up with it efficiently) was first proved in \cite{BMS13}:

\begin{lem}[Faithful is useful] \label{thm:faithful-pit} Let $\varphi$ be a homomorphism faithful to $\mathbf{f} = \{f_1,\ldots, f_m\} \subset \F[\mathbf{x}]$. Then for any $C\in\F[\mathbf{y}]$, $C(\mathbf{f}) = 0 \Leftrightarrow C(\varphi(\mathbf{f})) = 0$.
\end{lem}
This implies that we can use a faithful map to `reduce' the number of variables $n$ without changing the nonzeroness of $C$. The strategy can be used in cases where $\trdeg(\mathbf{f})$ is small, say, smaller than a constant $r$.

The only missing piece is the efficiency of $\varphi$\footnote{It can be shown, from first principles, that a faithful $r$-variate map always {\em exists} \cite{BMS13}.}. To do this we need three fundamental ingredients -- an efficient criterion for algebraic independence (Jacobian), its behavior under $\varphi$ (chain rule), and standard maps (Vandermonde \& Kronecker based).

\begin{lem}[Jacobian criterion] \label{fact:jacobi-criterion} Let $\mathbf{f} \subset \F[\mathbf{x}]$ be a finite set of polynomials of degree at most $d$, and $\trdeg(\mathbf{f}) \leq r$. If $\ch(\F) = 0$ or $\ch(\F) > d^r$, then $\trdeg(\mathbf{f}) = \rk_{\F(\mathbf{x})} \cJ_{\mathbf{x}}(\mathbf{f})$, where $\cJ_{\mathbf{x}}(\mathbf{f}) :=  \inparen{\partial f_i/ \partial x_j}_{m \times n}$ is the {\em Jacobian matrix}.
\end{lem}

There are several proofs of this, see \cite{J1841, For91, BMS13, MSS12}. 
This gives us an efficient way to capture trdeg, when the characteristic is zero/large. Let us now see how the Jacobian matrix changes under $\varphi$.

\begin{lem}[Chain rule]\label{fact:jacobian-chain-rule}
$\cJ_{\mathbf{y}}(\varphi(\mathbf{f})) = \varphi\inparen{\cJ_{\mathbf{x}}(\mathbf{f})} \cdot \cJ_{\mathbf{y}}(\varphi(\mathbf{x}))$, where $\varphi$ applied to a matrix/set refers to the matrix/set obtained by applying $\varphi$ to every entry.
\end{lem}

This is a simple consequence of the chain rule of `derivatives'. It suggests that for $\varphi$ to preserve the trdeg of the polynomials, we need to control -- (1) the image of the original Jacobian under $\varphi$, and (2) the Jacobian of the image of $\mathbf{x}$. In our applications, the former is achieved by a Kronecker-based map (i.e.~sparse PIT tricks, eg.~\cite{BHLV09}) and the latter by Vandermonde map (as seen in the previous subsection).

This general `recipe' has been successfully implemented to various circuit models. The case of the circuit $C'(\mathbf{x}) := C(\mathbf{f})$, where $\trdeg(\mathbf{f}) \leq r$ and $f_i$'s are polynomials of sparsity at most $s$, was worked out in \cite{BMS13}. The proof follows exactly the above strategy. The time complexity is polynomial in $\size(C')$ and $(s\cdot\degr(C'))^r$, where the exponential dependence comes from the sparsity estimate of $\cJ_{\mathbf{x}}(\mathbf{f})$ (and of course the final brute-force hitting-set for the $r$-variate $\varphi(C')$).

\cite{ASSS12} extended the recipe to depth-$4$ circuits (\ref{eqn-depth4}) where the number of $f_{i,j}$'s where any variable appears is bounded by $r$\footnote{Note that this does not mean that $\trdeg(f_{i,j}|i,j)$ is bounded.}. This model is called {\em occur-$r$} depth-$4$; it generalizes the well-studied multilinear read-$r$ depth-$4$. Interestingly, slightly modified techniques also provided {\em exponential} lower bounds against these special models. This required proving some combinatorial properties of the derivatives of immanant (eg.~permanent, determinant).

The faithful maps recipe has been able to unify all the assorted {\em poly-time} hitting-sets known. However, one drawback is that it needs the characteristic to be zero/large. Baby steps in resolving that issue have been taken by \cite{MSS12}.

\section{Rank concentration, shift, hitting-sets}

The hitting-sets that we saw till now were for models where some parameter was kept bounded. But we could also study models with a `structural' restriction, eg.~multilinearity. This route has also been successful and enlightening. We call a depth-$3$ circuit $C=\sum_i T_i$ {\em multilinear} if the linear factors in $T_i$ involve disjoint variables. Hence, each product gate $T_i$ induces a partition $\cP_i$ on the variables (or indices) $[n]$. Moreover, we call $C$ {\em set-multilinear} if these partitions are all equal!

There is a large body of work on the set-multilinear model \cite{RS05, AMS10, FS12, FS13a, ASS13, FS13b, FSS13, AGKS13}. The motivation for this model is, on the one hand, the algebraic concept of {\em tensors}, and, on the other hand, the interest in read-once {\em boolean} branching programs \cite{Nis92, IMZ12, Vad12}. Interestingly, \cite{FSS13} has shown (extending the ideas of \cite{ASS13}) that the situation in the arithmetic world is exponentially better than that in the boolean one!

Here we will exhibit the key ideas of \cite{ASS13} and \cite{AGKS13} on two {\em toy} cases that are already quite instructive; this saves us from the gory technical machinery that drives the more general cases.

\subsection{Multilinear ROABP}\label{sec-setMultilr}

\cite{ASS13} gave the first quasi-poly-time hitting-set for set-multilinear depth-$3$ (and extensions to constant-depth, non-multilinear versions). This was generalized by \cite{FSS13} to {\em any} depth; in fact, they dealt directly with the {\em multilinear ROABP} $D=\prod_i L_i$ over $M_w(\F)$, where $L_i$'s are linear polynomials in disjoint variables. Both the papers proved `low-support rank concentration' in their models. 

For the following discussion we fix a base commutative ring $R=H_w(\F)$ called the {\em Hadamard} algebra (instead of the $w\times w$ matrix algebra). This is basically $(\F^k,+,\star)$, where $+$ is the vector addition and $\star$ is the coordinate-wise vector product (called the Hadamard product). 

\smallskip\noindent{\bf $\ell$-concentration.} 
We say that a polynomial $f\in R[x_1,\ldots,x_n]$ is {\em $\ell$-concentrated} if 
$$\rk_\F \{\coeff_f (x_S) \mid S \subseteq [n], \abs{S} < \ell\} 
= \rk_\F \{\coeff_f (x_S) \mid S \subseteq [n]\},$$
where $\coeff_f$ extracts a coefficient in $f$.

I.e.~the coefficient-vectors of `lower' monomials already span every possible coefficient-vector in $f$. We are interested in studying whether circuits compute an $\ell$-concentrated polynomial for small $\ell$ (say, $\log n$ instead of $n$). By itself this is not true, eg.~the trivial circuit $D=x_1\cdots x_n$ is not even $n$-concentrated. But, maybe we can transform $f$ a bit and then attain $(\log n)$-concentration? In this case, $D':=D(x_1+1,\ldots,x_n+1)$ is suddenly $1$-concentrated!

It was shown by \cite{ASS13} that any $D$, above $R$, becomes $(\log k)$-concentra- ted after applying a `small' shift; the price of which is $n^{\log k}$ time. Once we have this it directly applies to the set-multilinear depth-$3$ model. Since, a depth-$3$ $C=\sum_{i\in[k]} T_i$ can be rewritten as $C=\insquare{1,\ldots,1}\cdot D$, where $D=\insquare{ \begin{array}{c}T_1\\\vdots\\T_k \end{array} }$ is of the promised sort over $R=H_k(\F)$ (since $D$ completely factorizes into disjoint-variate linear polynomials). So, $\ell$-concentration in $D$ implies an easy way to check $C$ for zeroness -- test the coefficients of the monomials below $\ell$-support in $C$.

\smallskip\noindent{\bf Glimpse of a proof.} 
We now show how to achieve $\ell$-concentration, $\ell=O(\log k)$, in the following toy model:
\begin{equation}\label{eqn-trivial-part}
D\ =\ \mathop{\prod}_{i\in[n]} (1+z_ix_i),\, \text{ where }z_i\in H_k(\F).
\end{equation}

Because of the disjointness of the factors it can be seen, as a simple exercise, that: $D$ is $\ell$-concentrated iff $D_S:=\prod_{i\in S} (1+z_ix_i)$ is $\ell$-concentrated, for all $S\in\binom{[n]}{\ell}$. Thus, from now on we assume, wlog, $n=\ell$.

Shift $D$ by formal variables $\mathbf{t}$, and normalize, to get a new circuit:
$$D'\ = \ \mathop{\prod}_{i\in[\ell]} (1+z'_ix_i),\, \text{ where }z'_i\in H_k(\F(\mathbf{t})).$$
We can express the new coefficients as:
$$z'_i = z_i/(1+z_it_i),\, \forall i\in[\ell].$$
Conversely, we write:
\begin{equation}\label{eqn-transfer}
z_i = z'_i/(1-z'_it_i),\, \forall i\in[\ell].
\end{equation}

We write $z_S$ for $\prod_{i\in S} z_i$. Now the goal is to `lift' an $\F$-dependence of $z_S$'s to the $z'_S$; which ultimately shows the condition on the shift that shall yield concentration.

Consider the $2^\ell$ vectors $\{z_S\; \mid\; S\subseteq[\ell]\}$. If $\ell>\log k$ then there is a nontrivial linear dependence amongst these vectors, say,
$$\mathop{\sum}_{S\subseteq[\ell]} \alpha_S z_S = 0,\, \text{ where } \alpha_S\in\F.$$

Rewriting this in terms of $z'_S$ we get:
\begin{eqnarray}
\mathop{\sum}_{S\subseteq[\ell]} \alpha_S\cdot 
\mathop{\prod}_{i\in S} z'_i/(1-z'_it_i) &=& 0. 
\nonumber\\
\text{Or, } \mathop{\sum}_{S\subseteq[\ell]} \alpha_S\cdot 
z'_S\cdot \mathop{\prod}_{i\in[\ell]\setminus S} (1-z'_it_i) &=& 0.
\label{eqn-lift}
\end{eqnarray}
 
Let us collect the `coefficient' of $z'_{[\ell]}$ in the above expression. It comes out to,
\begin{equation}\label{eqn-z-ell}
\mathop{\sum}_{S\subseteq[\ell]} \alpha_S\cdot 
(-1)^{\abs{[\ell]\setminus S}}\cdot t_{[\ell]\setminus S}.
\end{equation}

If we can ensure this expression to be nonzero then Equation (\ref{eqn-lift}) tells us that $z'_{[\ell]}$ is in the $\F(\mathbf{t})$-span of the `lower' $z'_S$. But, ensuring the nonzeroness of Equation (\ref{eqn-z-ell}) is easy -- use $t_i$'s such that all the $(\le\ell)$-support monomials $t_S$ are {\em distinct}. We can use standard sparse PIT tricks \cite{BHLV09} for this, in time $\poly(n^\ell)$. 

What we have shown is that, after applying a Kronecker-based shift, the circuit $D$ becomes $\ell$-concentrated; all this in time $n^{O(\log k)}$.
This `recipe' of studying the generic shift, via some combinatorial properties of the `transfer' equations (\ref{eqn-transfer}), is generalized in \cite{ASS13} to other $D$; and further improved in \cite{FSS13} to multilinear ROABP. It is not known how to design such hitting-sets, even for the toy case, in {\em poly}-time.

\subsection{Towards multilinear depth-$3$}

It is tantalizing to achieve $\ell$-concentration in multilinear depth-$3$ (before embarking on the general depth-$3$!). A partial result in that direction was obtained in \cite{AGKS13}. We will sketch their ideas in a toy model. 

Consider a multilinear depth-$3$ circuit $C$ with only {\em two} partitions being induced by the product gates -- $\cP_1=\inbrace{ \{1\}, \cdots, \{n\} }$ and an arbitrary partition $\cP_2$. Say, the number of the corresponding product gates is $k_1$ respectively $k_2$ (summing to $k$). We can say, naturally, that $\cP_1$ is a {\em refinement} of $\cP_2$ (denoted $\cP_1\le\cP_2$) because: For every color (or part) $S\in\cP_2$ there exist colors in $\cP_1$ whose union is {\em exactly} $S$. In this refinement situation \cite{AGKS13} showed that, again, a suitable shift in the $\prod\sum$ circuit $D$ (corresponding to $C$) achieves $\ell$-concentration in time $\poly(n^{\log k})$.

\smallskip\noindent{\bf Glimpse of a proof.}
We can assume $\cP_2$ different from $\cP_1$, otherwise this case is no different from the last subsection. We assume that the first $k_1$ product gates in $C=\sum_{i\in[k]} T_i$ respect $\cP_1$ and the rest $k_2$ respect $\cP_2$. The corresponding circuit $D$ where we desire to achieve concentration is $D=\insquare{ \begin{array}{c}T_1\\ \vdots \\T_k \end{array} }$ over $R=H_k(\F)$. But now the linear factors of $D$ are not necessarily in disjoint variables. Eg. 
$\insquare{ \begin{array}{c}x_1x_2\\ x_1+x_2 \end{array} } = 
\inparen{ x_1 + 
\insquare{ \begin{array}{c} 0\\1 \end{array} }\cdot x_2 
} \cdot 
\inparen{ \insquare{ \begin{array}{c} 0\\1 \end{array} } +
\insquare{ \begin{array}{c} 1\\0 \end{array} }\cdot x_2 
}$ over $H_2(\F)$.

To get some kind of a reduction to the set-multilinear case, we prove rank concentration in parts. First, we consider those monomials (called {\em $\cP_1$-type}) that could only be produced by the `upper' part of $D$ (i.e.~the first $k_1$ product gates of $C$). Such a monomial, say indexed by $S\subseteq[n]$, is characterized by the presence of $i,j\in S$ that are in the same color of $\cP_2$. For a fixed such $i, j$ we can ``access'' all such monomials by the derivative $\partial^2 D/\partial x_i\partial x_j =: \partial_{i,j}D$. Notice that this differentiation kills the `lower' part of $D$ and only the $\cP_1$-part remains. So, we can prove $(2+\log k_1)$-concentration in the monomials containing $i,j$ as in Section \ref{sec-setMultilr}. This proves $O(\log k_1)$-concentration in the monomials of $\cP_1$-type.

Next, we want to understand the remaining monomials (called $\cP_2$-type); those that could be produced by the `lower' part of $D$ (i.e.~the last $k_2$ product gates of $C$). These, obviously, could also be produced by the upper part of $D$. Let us fix such a monomial, say $x_1\cdots x_\ell$.
Assume that $S_1,\ldots,S_\ell\in\cP_2$ are the colors that contain one of the indices $1,\ldots,\ell$. Consider the subcircuit $D_{\ell}$ that in its $i$-th coordinate, $\forall i\in[k]$, simply drops those factors of $T_i$ that are free of the variables $S_1\cup\cdots\cup S_\ell$. The problem here is that $D_{\ell}$ may be a `high' degree circuit ($\approx n$ instead of $\ell$) and so we cannot use a proof like in Section \ref{sec-setMultilr}.  

But, notice that all the degree-$(\ge\ell)$ monomials in $D_{\ell}$ are $\cP_1$-type; where we know how to achieve $\ell$-concentration. So, we only have to care about degree-$(\le\ell)$ $\cP_2$-type monomials in $D_{\ell}$. There, again, $(\log k)$-concentration can be shown using Section \ref{sec-setMultilr} and the well-behaved transfer equations.

This sketch, handling two refined partitions, can be made to work for significantly generalized models \cite{AGKS13}. But, multilinear depth-$3$ PIT is still open (nothing better than exponential time known).

\begin{rem}
Using a different technique \cite{AGKS13} also proves {\em constant}-concentration, hence designs {\em poly}-time hitting-sets, for certain constant-width ROABP. These models are arithmetic analogues of the {\em boolean} ones -- width-$2$ read-once branching programs \cite{AGHP92, NN93} and constant-width read-once permutation branching programs \cite{KNP11}.  
\end{rem}

\section{Open ends}

The search for a strong enough technique to study arithmetic circuits continues. We collect here some easy-to-state questions that interest us.

\medskip\noindent{\bf Top fanin-$2$ depth-$4$.} Find a faithful map $\varphi$ that preserves the algebraic independence of two products-of-sparse polynomials $\prod_i f_i$ and $\prod_j g_j$. If we look at the relevant $2\times2$ Jacobian determinant, say wrt variables $X:=\{x_1, x_2\}$, then the question boils down to finding a hitting-set for the special {\em rational} function $\sum_{i,j}\frac{\det\cJ_X(f_i,g_j)}{f_ig_j}$. Can this version of {\em rational sparse} PIT be done in sub-exponential time?

\medskip\noindent{\bf Independence over $\F_p$.} Currently, there is no sub-exponential time algorithm/heuristic known to test two given circuits for algebraic independence over a `small' finite field $\F_p$. The reason is that something as efficient as the Jacobian criterion is not readily available, see \cite{MSS12}.  

\medskip\noindent{\bf Model in Eqn.(\ref{eqn-trivial-part}).} Find a {\em poly}-time hitting-set for this simple model. Note that a poly-time whitebox PIT is already known \cite{RS05}.

\medskip\noindent{\bf Multilinear depth-$3$.} Achieve $o(n)$-concentration in multilinear depth-$3$ circuits, in $n^{o(n)}$ time. Here, the presence of an exponential lower bound against the model \cite{RY09} is quite encouraging.

\bibliographystyle{amsalpha}
\bibliography{refs}

\providecommand{\bysame}{\leavevmode\hbox to3em{\hrulefill}\thinspace}
\providecommand{\MR}{\relax\ifhmode\unskip\space\fi MR }
\providecommand{\MRhref}[2]{%
  \href{http://www.ams.org/mathscinet-getitem?mr=#1}{#2}
}
\providecommand{\href}[2]{#2}
\begin{thebibliography}{AGHP92}

\bibitem[AGHP92]{AGHP92}
Noga Alon, Oded Goldreich, Johan H{\aa}stad, and Ren{\'e} Peralta, \emph{Simple
  construction of almost $k$-wise independent random variables}, Random Struct.
  Algorithms \textbf{3} (1992), no.~3, 289--304, (Conference version in FOCS
  1990).

\bibitem[AGKS13]{AGKS13}
Manindra Agrawal, Rohit Gurjar, Arpita Korwar, and Nitin Saxena,
  \emph{{Hitting-sets for low-distance multilinear depth-$3$}}, Electronic
  Colloquium on Computational Complexity (ECCC) \textbf{20} (2013), 174.

\bibitem[Agr05]{Agr05}
Manindra Agrawal, \emph{Proving lower bounds via pseudo-random generators},
  Proceedings of the 25th Annual Foundations of Software Technology and
  Theoretical Computer Science (FSTTCS), 2005, pp.~92--105.

\bibitem[Agr06]{Agr06}
\bysame, \emph{Determinant versus permanent}, Proceedings of the 25th
  {I}nternational {C}ongress of {M}athematicians (ICM), vol.~3, 2006,
  pp.~985--997.

\bibitem[AMS10]{AMS10}
Vikraman Arvind, Partha Mukhopadhyay, and Srikanth Srinivasan, \emph{{New
  Results on Noncommutative and Commutative Polynomial Identity Testing}},
  Computational Complexity \textbf{19} (2010), no.~4, 521--558, (Conference
  version in CCC 2008).

\bibitem[AS09]{AS09}
Manindra Agrawal and Ramprasad Saptharishi, \emph{{Classifying polynomials and
  identity testing}}, Indian Academy of Sciences, Platinum Jubilee \textbf{P1}
  (2009), 1--14.

\bibitem[ASS13]{ASS13}
Manindra Agrawal, Chandan Saha, and Nitin Saxena, \emph{{Quasi-polynomial
  hitting-set for set-depth-$\Delta$ formulas}}, STOC, 2013, pp.~321--330.

\bibitem[ASSS12]{ASSS12}
Manindra Agrawal, Chandan Saha, Ramprasad Saptharishi, and Nitin Saxena,
  \emph{{Jacobian hits circuits: hitting-sets, lower bounds for depth-$D$
  occur-$k$ formulas {\&} depth-$3$ transcendence degree-$k$ circuits}}, STOC,
  2012, pp.~599--614.

\bibitem[AV08]{AV08}
Manindra Agrawal and V.~Vinay, \emph{{Arithmetic circuits: A chasm at depth
  four}}, FOCS, 2008, pp.~67--75.

\bibitem[BC88]{BC88}
Michael {Ben-Or} and Richard Cleve, \emph{{Computing Algebraic Formulas Using a
  Constant Number of Registers}}, STOC, 1988, pp.~254--257.

\bibitem[BHLV09]{BHLV09}
Markus Bl\"aser, Moritz Hardt, Richard~J. Lipton, and Nisheeth~K. Vishnoi,
  \emph{Deterministically testing sparse polynomial identities of unbounded
  degree}, Inf. Process. Lett. \textbf{109} (2009), no.~3, 187--192.

\bibitem[BMS13]{BMS13}
Malte Beecken, Johannes Mittmann, and Nitin Saxena, \emph{Algebraic
  independence and blackbox identity testing}, Inf. Comput. \textbf{222}
  (2013), 2--19, (Conference version in ICALP 2011).

\bibitem[CKW11]{CKW11}
Xi~Chen, Neeraj Kayal, and Avi Wigderson, \emph{{Partial Derivatives in
  Arithmetic Complexity (and beyond)}}, Foundation and Trends in Theoretical
  Computer Science \textbf{6} (2011), no.~1-2, 1--138.

\bibitem[DS07]{DS07}
Zeev Dvir and Amir Shpilka, \emph{{Locally Decodable Codes with Two Queries and
  Polynomial Identity Testing for Depth 3 Circuits}}, SIAM J. Comput.
  \textbf{36} (2007), no.~5, 1404--1434, (Conference version in STOC 2005).

\bibitem[Fis94]{Fis94}
Ismor Fischer, \emph{Sums of like powers of multivariate linear forms},
  Mathematics Magazine \textbf{67} (1994), no.~1, 59--61.

\bibitem[For91]{For91}
Krister Forsman, \emph{Constructive commutative algebra in nonlinear control
  theory}, Ph.D. thesis, Dept. of Electrical Engg., Link\"oping University,
  Sweden, 1991.

\bibitem[FS12]{FS12}
Michael~A. Forbes and Amir Shpilka, \emph{On identity testing of tensors,
  low-rank recovery and compressed sensing}, STOC, 2012, pp.~163--172.

\bibitem[FS13a]{FS13b}
\bysame, \emph{{Explicit Noether Normalization for Simultaneous Conjugation via
  Polynomial Identity Testing}}, APPROX-RANDOM, 2013, pp.~527--542.

\bibitem[FS13b]{FS13a}
\bysame, \emph{{Quasipolynomial-time Identity Testing of Non-Commutative and
  Read-Once Oblivious Algebraic Branching Programs}}, FOCS, 2013.

\bibitem[FSS13]{FSS13}
Michael~A. Forbes, Ramprasad Saptharishi, and Amir Shpilka,
  \emph{Pseudorandomness for multilinear read-once algebraic branching
  programs, in any order}, Electronic Colloquium on Computational Complexity
  (ECCC) \textbf{20} (2013), 132.

\bibitem[GKKS13]{GKKS13}
Ankit Gupta, Pritish Kamath, Neeraj Kayal, and Ramprasad Saptharishi,
  \emph{Arithmetic circuits: A chasm at depth three}, FOCS, 2013.

\bibitem[GR08]{GR08}
Ariel Gabizon and Ran Raz, \emph{Deterministic extractors for affine sources
  over large fields}, Combinatorica \textbf{28} (2008), no.~4, 415--440,
  (Conference version in FOCS 2005).

\bibitem[HS80]{HS80}
Joos Heintz and Claus-Peter Schnorr, \emph{{Testing Polynomials which Are Easy
  to Compute (Extended Abstract)}}, STOC, 1980, pp.~262--272.

\bibitem[IMZ12]{IMZ12}
Russell Impagliazzo, Raghu Meka, and David Zuckerman, \emph{Pseudorandomness
  from shrinkage}, FOCS, 2012, pp.~111--119.

\bibitem[Jac41]{J1841}
Carl Gustav~Jacob Jacobi, \emph{De determinantibus functionalibus}, J. Reine
  Angew. Math. \textbf{22} (1841), no.~4, 319--359.

\bibitem[KI04]{KI04}
Valentine Kabanets and Russell Impagliazzo, \emph{{Derandomizing Polynomial
  Identity Tests Means Proving Circuit Lower Bounds}}, Computational Complexity
  \textbf{13} (2004), no.~1-2, 1--46, (Conference version in STOC 2003).

\bibitem[KNP11]{KNP11}
Michal Kouck{\'y}, Prajakta Nimbhorkar, and Pavel Pudl{\'a}k,
  \emph{Pseudorandom generators for group products: extended abstract}, STOC,
  2011, pp.~263--272.

\bibitem[Koi12]{Koi12}
Pascal Koiran, \emph{{Arithmetic circuits: The chasm at depth four gets
  wider}}, Theor. Comput. Sci. \textbf{448} (2012), 56--65.

\bibitem[KS07]{KS07}
Neeraj Kayal and Nitin Saxena, \emph{{P}olynomial {I}dentity {T}esting for
  {D}epth 3 {C}ircuits}, Computational Complexity \textbf{16} (2007), no.~2,
  115--138, (Conference version in CCC 2006).

\bibitem[KS09]{KS09}
Neeraj Kayal and Shubhangi Saraf, \emph{{Blackbox polynomial identity testing
  for depth-$3$ circuits}}, FOCS, 2009, pp.~198--207.

\bibitem[KS11]{KS11}
Zohar~Shay Karnin and Amir Shpilka, \emph{Black box polynomial identity testing
  of generalized depth-3 arithmetic circuits with bounded top fan-in},
  Combinatorica \textbf{31} (2011), no.~3, 333--364, (Conference version in CCC
  2008).

\bibitem[KSS13]{KSS13}
Neeraj Kayal, Chandan Saha, and Ramprasad Saptharishi, \emph{A super-polynomial
  lower bound for regular arithmetic formulas}, Electronic Colloquium on
  Computational Complexity (ECCC) \textbf{20} (2013), 91.

\bibitem[Mit13]{Mit13}
Johannes Mittmann, \emph{{Independence in Algebraic Complexity Theory}}, Ph.D.
  thesis, Mathematisch-Naturwissenschaftlichen Fakult\"at der Rheinischen
  Friedrich-Wilhelms-Universit\"at Bonn, Germany, December 2013.

\bibitem[MSS12]{MSS12}
Johannes Mittmann, Nitin Saxena, and Peter Scheiblechner, \emph{{Algebraic
  Independence in Positive Characteristic -- A $p$-adic Calculus}}, Electronic
  Colloquium on Computational Complexity \textbf{TR12-014} (2012), (accepted in
  Trans. Amer. Math. Soc., 2013).

\bibitem[Mul11]{Mul11}
Ketan Mulmuley, \emph{{On P vs.~NP and geometric complexity theory: Dedicated
  to Sri Ramakrishna}}, J. ACM \textbf{58} (2011), no.~2, 5.

\bibitem[Mul12a]{Mul12}
\bysame, \emph{{Geometric Complexity Theory V: Equivalence between Blackbox
  Derandomization of Polynomial Identity Testing and Derandomization of
  Noether's Normalization Lemma}}, FOCS, 2012, pp.~629--638.

\bibitem[Mul12b]{Mul-cacm12}
\bysame, \emph{{The GCT program toward the P vs.~NP problem}}, Commun.~ACM
  \textbf{55} (2012), no.~6, 98--107.

\bibitem[MV97]{MV97}
Meena Mahajan and V.~Vinay, \emph{{Determinant: Combinatorics, Algorithms, and
  Complexity}}, Chicago J. Theor. Comput. Sci. (1997), (Conference version in
  SODA 1997).

\bibitem[Nis92]{Nis92}
Noam Nisan, \emph{Pseudorandom generators for space-bounded computation},
  Combinatorica \textbf{12} (1992), no.~4, 449--461, (Conference version in
  STOC 1990).

\bibitem[NN93]{NN93}
Joseph Naor and Moni Naor, \emph{{Small-Bias Probability Spaces: Efficient
  Constructions and Applications}}, SIAM J. Comput. \textbf{22} (1993), no.~4,
  838--856, (Conference version in STOC 1990).

\bibitem[RS05]{RS05}
Ran Raz and Amir Shpilka, \emph{Deterministic polynomial identity testing in
  non-commutative models}, Computational Complexity \textbf{14} (2005), no.~1,
  1--19, (Conference version in CCC 2004).

\bibitem[RY09]{RY09}
Ran Raz and Amir Yehudayoff, \emph{Lower bounds and separations for constant
  depth multilinear circuits}, Computational Complexity \textbf{18} (2009),
  no.~2, 171--207, (Conference version in CCC 2008).

\bibitem[Sap13]{Sap13}
Ramprasad Saptharishi, \emph{{Unified Approaches to Polynomial Identity Testing
  and Lower Bounds}}, Ph.D. thesis, Department of CSE, IIT Kanpur, India, April
  2013.

\bibitem[Sax08]{Sax08}
Nitin Saxena, \emph{{Diagonal Circuit Identity Testing and Lower Bounds}},
  ICALP (1), 2008, pp.~60--71.

\bibitem[Sax09]{Sax09}
\bysame, \emph{{Progress on Polynomial Identity Testing}}, Bulletin of the
  EATCS (2009), no.~90, 49--79.

\bibitem[SS11]{SS11}
Nitin Saxena and C.~Seshadhri, \emph{{An Almost Optimal Rank Bound for Depth-3
  Identities}}, SIAM J. Comput. \textbf{40} (2011), no.~1, 200--224,
  (Conference version in CCC 2009).

\bibitem[SS12]{SS12}
\bysame, \emph{Blackbox identity testing for bounded top-fanin depth-3
  circuits: The field doesn't matter}, SIAM J. Comput. \textbf{41} (2012),
  no.~5, 1285--1298, (Conference version in STOC 2011).

\bibitem[SS13]{SS13}
\bysame, \emph{{From Sylvester-Gallai configurations to rank bounds: Improved
  blackbox identity test for depth-3 circuits}}, J. ACM \textbf{60} (2013),
  no.~5, 33, (Conference version in STOC 2010).

\bibitem[SSS09]{SSS09}
Chandan Saha, Ramprasad Saptharishi, and Nitin Saxena, \emph{{The Power of
  Depth 2 Circuits over Algebras}}, FSTTCS, 2009, pp.~371--382.

\bibitem[SY10]{SY10}
Amir Shpilka and Amir Yehudayoff, \emph{{Arithmetic Circuits: A survey of
  recent results and open questions}}, Foundations and Trends in Theoretical
  Computer Science \textbf{5} (2010), no.~3-4, 207--388.

\bibitem[Tav13]{Tav13}
S\'ebastien Tavenas, \emph{{Improved Bounds for Reduction to Depth 4 and Depth
  3}}, MFCS, 2013, pp.~813--824.

\bibitem[Vad12]{Vad12}
Salil~P. Vadhan, \emph{Pseudorandomness}, Foundations and Trends in Theoretical
  Computer Science \textbf{7} (2012), no.~1-3, 1--336.

\bibitem[Val79]{V79}
Leslie~G. Valiant, \emph{Completeness classes in algebra}, STOC, 1979,
  pp.~249--261.

\bibitem[VSBR83]{VSBR83}
Leslie~G. Valiant, Sven Skyum, Stuart~J. Berkowitz, and Charles Rackoff,
  \emph{{Fast Parallel Computation of Polynomials Using Few Processors}}, SIAM
  J. Comput. \textbf{12} (1983), no.~4, 641--644.

\bibitem[vzG83]{G83}
Joachim von~zur Gathen, \emph{{Factoring Sparse Multivariate Polynomials}},
  FOCS, 1983, pp.~172--179.

\end{thebibliography}

\end{document}